\documentclass{article}
\begin{document}
\pagestyle{empty}

\title{Simple Optimal Wait-free Multireader Registers\\
}


\author{Paul Vit\'anyi
\thanks{Centrum voor Wiskunde en Informatica (CWI),
Kruislaan 413, 1098 SJ Amsterdam,
The Netherlands.
Email: paulv@cwi.nl.
This work was supported in part
by the EU fifth framework project QAIP, IST--1999--11234,
the NoE QUIPROCONE IST--1999--29064, the ESF QiT Programmme,
and the EU Fourth Framework BRA NeuroCOLT II Working Group EP 27150.
}
}
\maketitle
\newcommand{\ra}{\longrightarrow}
\newcommand{\LA}{\Longrightarrow}
\newcommand{\RA}[1]{\Longrightarrow_{#1}}
\newcommand{\cf}{\stackrel{cf}{\ra}}
\newcommand{\da}{\parbox[c]{22pt}{\begin{picture}(22,0)(0,0) \multiput(4,0)(5,0){2}{\line(1,0){3}}
                                                  \put(14,0){\vector(1,0){4}}
                       \end{picture}}}

\newtheorem{theorem}{Theorem}
\newtheorem{lemma}{Lemma}
\newtheorem{assertion}{Assertion}
\newtheorem{claim}{Claim}
\newtheorem{corollary}{Corollary}
\newtheorem{proposition}{Proposition}
\newtheorem{definition}{Definition}

\newcommand{\ar}{\prec}
\newcommand{\arq}{\preceq}
\newcommand{\Ar}{<}
\newcommand{\qc}{\mbox{\rm c}}
\newcommand{\qs}{\mbox{\rm s}}
\newcommand{\qh}{\mbox{\rm h}}
\newcommand{\qr}{\mbox{\rm r}}
\newcommand{\qt}{\mbox{\rm t}}
\newcommand{\qp}{\mbox{\rm p}}
\newcommand{\qw}{\mbox{\rm w}}
\newcommand{\pLast}{\mbox{\rm p\it -Last}}
\newcommand{\nseq}{0..n}
\newcommand{\nseqw}{0..2n-2}
\newcommand{\abs}{\mbox{{\sc abs}}}
\def\?#1?{\mbox{\it #1\/}}
\def\|#1|{\mbox{\bf #1\/}}
\newcommand{\assign}{:=}
\newcommand{\increment}{+:=}
\newcommand{\mincrement}{\oplus :=}
\newenvironment{proof} {\vspace{0.01cm} {\bf Proof}.  }{ \hfill $\bullet$}

\newcommand{\buff}{\mbox{{\em buff\/}}}
\newcommand{\RC}{\mbox{{\em r\/}}}
\newcommand{\WC}{\mbox{{\em w\/}}}
\newcommand{\seen}{\mbox{{\em seen\/}}}
\newcommand{\copyc}{\mbox{{\em copyc\/}}}
\newcommand{\copybuff}{\mbox{{\em copybuff\/}}}
\newcommand{\prevc}{\mbox{{\em prevc\/}}}

\newcommand{\wpk}{\mbox{$W_{p}^{[k]}$}}
\newcommand{\wpki}{\mbox{$W_{p}^{[k+1]}$}}
\newcommand{\wpkj}{\mbox{$W_{p}^{[k+2]}$}}
\newcommand{\wqt}{\mbox{$W_{q}^{[t]}$}}
\newcommand{\wqti}{\mbox{$W_{q}^{[t+1]}$}}
\newcommand{\wqtj}{\mbox{$W_{q}^{[t+2]}$}}
\newcommand{\ril}{\mbox{$R_{i,p}^{[l]}$}}
\newcommand{\rilj}{\mbox{$R_{i,j}^{[l]}$}}
\newcommand{\rili}{\mbox{$R_{i}^{[l+1]}$}}
\newcommand{\rpi}{\mbox{$r_{i,p}$}}
\newcommand{\rqj}{\mbox{$r_{q,j}$}}

\newcommand{\wpi}{\mbox{$w_{p,i}$}}
\newcommand{\wqj}{\mbox{$w_{q,j}$}}

\newcommand{\wpj}{\mbox{$w_{p,j}$}}

\newcommand{\rjm}{\mbox{$R_{j}^{[m]}$}}

\renewcommand{\index}{\mbox{$index$}}
\begin{abstract}
Multireader shared registers are basic objects used as communication medium
in asynchronous concurrent computation. 
We propose a surprisingly simple and natural scheme 
to obtain several wait-free constructions of bounded  
1-writer multireader registers from atomic 1-writer 1-reader registers,
that is easier to prove correct than any previous construction.
Our main construction is the first symmetric pure timestamp one that is
optimal with respect to the worst-case local use of control
bits; the other one is optimal with respect to global use of
control bits; both are optimal in time. 
\end{abstract}


%

\section{Introduction} \label{sec1}
Interprocess communication in distributed systems happens by
either message-passing or shared variables (also known
as shared {\em registers}). Lamport \cite{lamp86}
argues that every message-passing system must hide an underlying
shared variable. This makes the latter an indispensable ingredient.
The multireader  wait-free shared variable is
(even more so than the multiwriter construction)
the building block of virtually all bounded wait-free shared-variable
concurrent object constructions, for example, 
\cite{vit86,isra93,dole89,gawl92,dwork92,dwork92a,HV01,abra88,isra92,pete87}.
Hence, understanding, simplicity, and optimality of bounded wait-free
atomic  multireader
constructions is a basic concern. 
Our 
constructions are really simple and structured extensions
of the basic unbounded timestamp algorithm in \cite{vit86},
based on a natural recycling scheme of obsolete timestamps.

{\bf Asynchronous communication:}
Consider a system of asynchronous processes that communicate among themselves
by executing read and write operations on a set of 
shared variables 
only.
The system has no global clock or other synchronization primitives.
Every shared variable is associated with a process (called {\em owner\/}) which writes it and
the other processes may read it.
An execution of a write (read) operation on a shared variable will
be referred to as a {\em Write\/}  ({\em  Read\/}) on that variable.
A Write on a shared variable puts a value from a pre determined finite domain into the
variable, and a Read   reports a value from the domain.
A process that writes (reads) a variable is called a {\em writer\/} ({\em reader\/}) of the
variable.

{\bf Wait-free shared variable:}
We want to construct shared variables in which 
the following two properties hold.
(1)~Operation executions are not necessarily atomic, 
that is, they are not indivisible,
and (2)~every operation finishes its execution 
within a bounded number of its
own steps, irrespective of the presence of other operation executions and
their relative speeds. That is, operation executions are {\em wait-free}.
These   two properties give rise to a classification of shared variables:
(i) A {\em safe\/} variable is one in which a Read not overlapping any Write
returns the most recently written value.
A Read that overlaps a Write may return any value from the domain of the variable;
(ii)  A {\em regular\/} variable is a safe variable in which
a Read that overlaps one or more Writes returns either the value
of the most recent Write preceding the Read or of one of the overlapping
Writes; and
%
(iii) An {\em atomic\/} variable is a regular variable in which
the Reads and Writes behave as if they occur in some total order
which is an extension of the precedence relation.
 

 {\bf Multireader shared variable:}
A multireader shared variable is one that can be written by one process 
and read
(concurrently) by many processes.
Lamport  \cite{lamp86} constructed an atomic wait-free shared variable
that could be written by one process and read by one other
process, but he did not consider constructions
of wait-free shared variables with more than one writer or reader.

{\bf Previous Work:}
In 1987 there appeared at least five purported solutions for the
wait-free implementation of 1-writer $n$-reader atomic shared variable from
atomic 1-writer 1-reader shared variables: 
\cite{kiro87,wolf87,burn87,sing87} and
the conference version of \cite{isra93},
of which \cite{burn87} was shown to be incorrect in \cite{hald92}
and only \cite{sing87} appeared in journal version.
The only other 1-writer $n$-reader atomic shared variable constructions
appearing in journal version are \cite{hald91} and the ``projection''
of the main construction in \cite{li89}.
A selection of related work is
\cite{vit86,li89,bloom87,burn87,hald91,hald95,hald96,isra92a,kiro87,lamp86,LV92,li89,pete83,pete87,scha88,sing87,wolf87}.
A brief history of atomic wait-free shared variable
constructions and timestamp systems is given in \cite{HV01}. 


Israeli and Li \cite{isra93} introduced
and analyzed the notion of
{\em timestamp system\/} as an abstraction of 
a higher typed communication medium than shared variables. 
(Other constructions are
\cite{dole89,isra92,gawl92,dwork92,dwork92a,HV01}.)
As an example of its application, \cite{isra93} presented
a non-optimal multireader construction, partially based on \cite{vit86},
using order $n^3$ control bits overall, and order $n$ control bits locally
for each of $O(n^2)$ shared 1-reader 1-writer variables.
Our constructions below are inspired by this timestamp method 
and exploit the limited nature of the multireader
problem to obtain both simplification and optimality.

 {\bf This work:}
We give a surprisingly simple and elegant wait-free construction of
an atomic multireader shared variable from atomic 1-reader 1-writer
shared variables.
Other constructions (and our second construction)
don't use pure timestamps \cite{isra93} but divide the timestamp
information between the writer's control bits and the reader's control
bits. Those algorithms have
an asymmetric burden on the writer-owned subvariables. Our construction shows
that using pure timestamps we can balance the size of 
the subvariables evenly among all owners,
obtaining a symmetric construction. The intuition behind our
approach is given in Section~\ref{sect.intuition}. Technically: 

(i) Our construction
is a bounded version of the unbounded construction in \cite{vit86}
restricted to the multireader case (but different from the ``projection''
of the bounded version in \cite{li89}), and its correctness proof follows
simply and directly from the correctness 
of the unbounded version and a proof of 
proper recycling of obsolete timestamps. 

(ii) The main version 
of our construction is the first construction that uses a sublinear number
of control
bits, $O(\log n)$, locally {\em for  everyone} of the $n^2$ 
constituent 1-reader 1-writer subvariables. It is easy to see that
this is optimal locally, leading to
a slightly non-optimal global number
of control bits of order $n^2 \log n$. Implementations of wait-free
shared variables assume atomic shared (control) subvariables. Technically 
it may be much simpler to manufacture in hardware---or implement
in software---small $\Theta (\log n)$-bit
atomic subvariables that are robust and reliable, than the exponentially
larger $\Theta (n)$-bit atomic 
subvariables required in all previous constructions.
A wait-free implementation of a shared variable is
fault-tolerant against crash failures of the subvariables, but not
against more malicious, say Byzantine, errors. Slight suboptimality
is a small price to pay for ultimate reliability.

(iii) Another version of our
construction uses $n$ 1-writer 1-reader variables of $n$ control
bits and $n^2$ 1-writer 1-reader variables of $2$ control bits each,
yielding a global $O(n^2)$ number of control bits.
There need to be $n ^ 2 $ 1-reader 1-writer subvariables for
pairwise communication, each of them
using some control bits. Hence the global number of control bits
in any construction is $\Omega (n^2)$.
We also reduce the number of copies of the
value written to $O(n)$ rather than $O(n^2)$.
All these variations of our construction
use the minimum number $O(n)$ of accesses to 
the shared 1-reader 1-writer subvariables, per Read and Write 
operation.

\begin{table}[h]
\center{
\begin{tabular}{|c|c|c|c|} \hline
Construction&Global control bits& Max local control bits  \\ \hline
\cite{kiro87} & $\Theta(n^3)$ & $\Theta(n)$   \\ \hline
\cite{wolf87} & $\Theta(n^2)$ & $\Theta(n)$   \\ \hline
\cite{burn87}-incorrect: \cite{hald92} & $\Theta(n^2)$ & $\Theta(n)$   \\ \hline
\cite{sing87} & $\Theta(n^2)$ & $\Theta (n)$   \\ \hline
\cite{isra93} & $\Theta (n^3)$ & $\Theta (n)$   \\ \hline
\cite{li89} & $\Theta (n^2)$ & $\Theta (n)$   \\ \hline
\cite{hald91} & $\Theta (n^2)$ & $\Theta (n)$   \\ \hline
This paper (ii) & $\Theta (n^2 \log n)$ & $\Theta (\log n)$  \\ \hline
This paper (iii) & $\Theta (n^2)$ & $\Theta (n)$   \\ \hline
\end{tabular}
\caption{Comparison of Results.}\label{tab1}
}
\end{table}
Our construction corrects \cite{Vi02} by merging pairs $R_{i,n}, R_{i,n+1}$ of
shared variables in that paper into a single shared variable $R_{i,n}$ here
($0 \leq i < n$). We give an informal argument
why the construction works correctly; the formal argument is deferred
to a later treatment in I/O automata assertional framework surveyed in
\cite{KLSV04}.

\subsection{Model, Problem Definition, and some Notations}\label{sec2}
\label{constr}
Throughout the paper, there are $n$ readers indexed $0, 1, \ldots , n-1$
and a single writer
indexed $n$. 
The multireader variable constructed will be called \abs\ (for
abstract).

A construction consists of a collection of atomic 1-reader 1-writer
shared variables $R_{i,j}$,
$i,j \in \{0, 1, \ldots , n\}$, written by process $i$ and read by process $j$,
thus providing a communication path from user $i$ to user $j$. 
We say that the writer, process $i$, {\em owns} the shared
variable $R_{i,j}$.
Stricly speaking, the ``diagonal'' shared  
variables $R_{i,i}$ ($0 \leq i \leq n$) are superfluous since process $i$
can remember in a local variable what it wrote last to a shared variable
it owns. We keep the diagonal shared variables nonetheless,
 since doing so allows us to
simplify the protocols.
There are two global procedures, \?Read? and \?Write?, to execute a read operation
or write operation on \abs\ , respectively. Each of these are expressed in terms
of multiple reads and writes on the shared variables involved. 
Both procedures have an input parameter $i$, which is the index of the
executing user, and in addition,
Write takes a value to be written to \abs\ as input.
A return statement must end both procedures,
in the case of Read having an argument which is
taken to be the value read from \abs.

A procedure contains a declaration of local variables and a body.
A local variable appearing in can be declared {\em static},
which means it retains its value between procedure invocations.
The body is a program fragment comprised of atomic statements.
Access to shared variables is naturally restricted to
assignments from $R_{j,i}$ to local variables and assignments from local
variables to $R_{i,j}$, for any $j$ (recall that $i$ is the index of
the executing user). No other means of inter-process communication is allowed.
In particular, no synchronization primitives can be used.
Assignments to and from shared variables are called writes and reads
respectively, always in lower case.
The {\em space complexity} of a construction is the maximum size, in bits,
of a shared variable.
The {\em time complexity} of the Read or Write procedure
is the maximum number of
shared variable accesses in a single execution.

\subsection{Correctness}
A wait-free construction must satisfy the following constraint:
{\bf Wait-Freedom:} Each procedure must be free from unbounded loops.
Given a construction, we are interested in properties of its executions,
which the following notions help formulate.
A {\em state} is a configuration of the construction, comprising values
of all shared and local variables, as well as program counters.
In between invocations
of the Read and Write procedure, a user is said to be {\em idle}, and its
program counter has the value `idle'.
One state is designated as {\em initial state}.
All users must be idle in this state.

A state $t$ is an {\em immediate successor} of a state $s$ if $t$ can
be reached from $s$ through the execution of a procedure statement by
some user in accordance with its program counter.
Recall that $n$ denotes the number of readers of the constructed variable \abs.
A state has precisely $n+1$ immediate successors: 
there is at precisely one
atomic statement per process to be executed next (each process is
deterministic).

A {\em history} of the construction is a finite or infinite sequence of
states $t_0,t_1,\ldots$ such that $t_0$ is the initial state and
$t_{i+1}$ is an immediate successor of $t_i$.
Transitions between successive states are called the {\em events} of
a history. With each event is associated the index of the executing user,
the relevant procedure statement, and the values manipulated by
the execution of the statement. Each particular access to a shared variable
is an event, and all such events are totally ordered.

The {\em (sequential) time complexity} of the Read or Write procedure
is the maximum number of shared variable accesses in some such operation in
some history.

An event $a$ {\em precedes} an event $b$ in history $h$, $a \ar_h b$,
if $a$ occurs before $b$ in $h$. 
Call a finite set of events of a history an event-set.
Then we similarly
say that an event-set $A$ precedes an event-set $B$
in a history, $A \ar_h B$,
when each event in $A$ precedes all those in $B$.
We use $a \arq_h b$ to denote that either $a=_hb$ or $a \ar_h b$.
The relation $\ar_h$ on event-sets constitutes what is known as an
{\em interval order}. That is, a partial order $\ar$ satisfying the interval
axiom $a \ar b \wedge c \ar d \wedge c \not\ar b \Rightarrow a \ar d$.
This implication can be seen to hold by considering the last event of $c$ and
the earliest event of $b$. See \cite{lamp86} for an extensive discussion on
models of time.

Of particular interest are the sets consisting
of all events of a single procedure invocation, which we call an
{\em operation}. An operation is either a Read operation or a Write operation.
It is {\em complete} if it includes the execution of the final {\bf return}
statement of the procedure. Otherwise it is said to be {\em pending}.
A history is complete if all its operations are complete. Note that in the
final state of a complete finite history, all users are idle.
The {\em value} of an operation is the value written to \abs\ in the case
of a Write, or the value read from \abs\ in the case of a Read.

The following crucial definition expresses the idea that the operations
in a history appear to take place instantaneously somewhere during their
execution interval. A more general version of this is presented and motivated
in \cite{herl90}.
To avoid special cases, we introduce the notion of a {\em proper} history
as one that starts with an initializing Write operation that precedes all
other operations.
{\bf Linearizability:}
A complete proper history $h$ is {\em linearizable} if the
partial order $\ar_h$ on the set of operations can be extended
to a total order which obeys the semantics of a variable. That is,
each Read operation returns the value written by that Write operation
which last precedes it in the total order.
We use the following definition and lemma from \cite{li89}:

\begin{definition}
A construction is {\em correct} if it satisfies Wait-Freedom
and all its complete proper histories are linearizable.
\end{definition}

\begin{lemma}
A complete proper history $h$ is linearizable iff there exists a function
mapping each operation in $h$ to a rational number, called its timestamp,
such that the following 3 conditions are satisfied:

{\bf Uniqueness:}
different Write operations have different timestamps.

{\bf Integrity:}
for each Read operation there exists a Write operation
with the same timestamp and value, that it doesn't precede.

{\bf Precedence:} if one operation precedes another, then the timestamp
of the latter is at least that of the former.
\label{taglemma}
\end{lemma}

\section{Intuition}
\label{sect.intuition}
We use the following intuition based on
timestamp systems:
 The concurrent reading and writing
of the shared variable by one writer and $n$ readers 
(collectively, the {\em players}) is viewed
as a pebble game on a finite directed graph. The nodes of the graph 
can be viewed as the timestamps used by the system, and a pebble on a node
as a subvariable containing this timestamp. If the graph
contains an arc pointing from node $a$ to node $b$ then $a$
is {\em dominated} by $b$ ($a < b$). The graph has no cycles of length 1 or 2.
First, suppose that every player has a single pebble
which initially is placed at a distinguished node that is dominated
by all other nodes. A Read or Write by a player
consists in observing where the pebbles of the other players
are, and determining a node to move its own pebble to.
In a Write, the writer puts its pebble on
a node that satisfies all of the following: 
(i) it dominates the previous position; 
(ii) it is not occupied
by a pebble of a reader; and 
(iii) it is not dominated
by a node occupied by a pebble of a reader. In a Read, the reader
concerned puts its pebble at a node containing the writer's pebble.
In the pebble game a player can only observe the pebble
positions of the other players in sequence, one at a time.
By the time the observation sequence is finished, pebbles observed
in the beginning may have moved again. Thus, in an observation sequence
by a reader, a node can contain another reader's pebble that dominates
the node containing the writer's pebble. Then,
the second reader has already observed the node pebbled by
a later Write of the writer
and moved its pebble there. In the unbounded timestamp
solution below, where the timestamps are simply the nonnegative integers
and a higher integer dominates a lower integer,
this presents no problem: a reader simply moves its
pebble to the observed pebbled node corresponding to the greatest timestamp.
But in the bounded timestamp solution we must distinguish 
obsolete timestamps from active recycled timestamps.
Clearly, the above scenario cannot happen
if in a Read we observe
the position of the writer's pebble last in the obervation sequence. 
 Then, the linearization of the
complete proper history of the system consists of ordering
every Read after the Write it joined its pebble with.
This is the basic idea. But there is a complication that makes life
less easy.

In the implementation of the 1-writer $n$-reader variable in 1-writer
1-reader subvariables, every subvariable is a one-way communication
medium between two processes. Since two-way communication
between every pair of processes is required, there are at least
$(n+1)n$ such subvariables: Every process owns at least $n$
subvariables it can write, each of which is read by one of
the other $n$ processes. Thus, a pebble move of a player
actually consists in moving (at least) $n$ pebble copies,
in sequence one at a time---a move of a pebble 
consists in atomically writing
a subvariable. Every pebble copy can only
be observed by a single fixed other player---the reader of 
that 1-writer 1-reader
subvariable. This way player 0 may move the pebble copy associated
with player 1 to another node (being observed by player 1 at time $t_1$)
while the pebble copy associated with player 2 is still at the originating
node (being observed later by player 2 at time $t_2 > t_1$).
This once more opens the possibility that a reader sees a writer's pebble at
a node pebbled by a Write, 
while another reader earlier on sees a writer's pebble
at a node pebbled by a later Write. The following argument shows that
this `later Write' must be in fact the `next Write':
A reader always joins its pebble to a node that
is already pebbled, and only the writer can move its pebbles to an unoccupied
node. Therefore, the fact that a reader observes the writer's pebble last
guaranties that no reader, of which the pebble was observed
before, can have observed a Write's pebble position later than the next Write. 
Namely, at the time the writer's pebble was observed by the former
reader at the node pebbled by Write, the next Write wasn't finished, but
the latter reader has already observed the writer's pebble.
This fact is important in the bounded timestamp solution we present
below, since it allows us to use a very small timestamp graph
that contains cycles of length just 3: $G=(V,E)$ with
$V = \{ \perp, 1, \ldots, 4n+3\}^2$ and $(v_1,v_2) \in E$ ($v_1 < v_2$)
iff either $v_1 = (i,j), v_2 = (k,h)$ and $j=k$ and $i \neq h \neq \perp$,
 or $(i,j) = (\perp,\perp)$
and $h \neq \perp$. 

This approach suffices to linearize the complete proper history
of the system if we can prevent a Write to pebble the same node
as a Read in the process of chasing an older obsolete
Write. This we accomplish by a Read by player $i$
announcing its intended destination node
several times to the writer, with an auxiliary pebble for the
purpose (special auxiliary subvariable)
intended for this purpose, and in between checking whether the writer has 
moved to the same node. In this process either the Read discovers
a Write that is intermediate between two Writes it observed,
and hence covered by the Read, in which case it can safely report
that Write and choose the bottom timestamp $(\perp, \perp)$.
Alternatively, the Read ascertains that future Writes know
about the timestamp it intends to use and those Writes will avoid that
timestamp. 

\section{Unbounded}
\label{sect.unbounded}

Figure \ref{cons0} shows Construction~0, which is a restriction
to the multireader case of the unbounded solution multiwriter construction
of \cite{vit86}. Line 2 of the Write
procedure has the same effect as ``$\?free? \assign \?free?+1$'' 
with $\?free?$ initialized at 0 (because the writer always writes
$\?free?$ to $R_{n,n}.\?tag?$ in line 4). 
The processes indexed $0, \ldots, n-1$
are the readers and the process indexed $n$ is the writer.
We present it here as an aid in understanding Construction~1.
Detailed proofs of correctness (of the
unrestricted version, where not just process $n$ can Write,
but every process can do so)
are given in \cite{vit86} and essentially simple
proofs in  \cite{li89} and the textbook \cite{Ly97}.


\begin{figure}[h]
\fbox{%
\begin{minipage}[t]{3in}
\begin{tabbing}
\|type| \= \?shared? : \=  \|re| \= \kill
\|type| \= $I : 0..n$ \\
        \> \?shared? : \|record| \\
        \>             \>    \>\?value? : \abs\?type?  \\
        \>             \>    \>\?tag? \ : \|integer| \\
        \>             \> \|end|
\end{tabbing}
\end{minipage}
}

\medskip

\fbox{%
\begin{minipage}[t]{3in}
\begin{tabbing}
\|procedure| $\?Write?(n,v)$ \\
\|var| \= $j : I$ \\
       \> $\?free? :$ \|integer| \\
       \> \?from? : $\|array|[\nseq]$ \|of| \?shared? \\
\|begin| \\
1        \> \|for| $j \assign 0 .. n$ \|do| $\?from?[j] \assign R_{j,n}$ \\
2        \> $\?free? \assign \max_{j \in I} \?from?[j].\?tag? + 1$ \\
3        \> $\?from?[n] \assign (v, \?free?)$ \\
4        \> \|for| $j \assign 0 .. n$ \|do| $R_{n,j} \assign \?from?[n]$ \\
\|end|
\end{tabbing}
\end{minipage}
}

\medskip

\fbox{
\begin{minipage}[t]{3in}
\begin{tabbing}
\|procedure| $\?Read?(i)$ \\
\|var| \= $j, \?max? : I$ \\
       \> $\?from? : \|array|[\nseq]$ \|of| \?shared? \\
\|begin| \\
1        \> \|for| $j \assign 0 .. n$ \|do| $\?from?[j] \assign R_{j,i}$ \\
2        \> select \?max? such that $\forall j : \?from?[\?max?].\?tag?
           \geq \?from?[j].\?tag?$ \\
3        \> $\?from?[i] \assign \?from?[\?max?]$ \\
4        \> \|for| $j \assign 0 .. n$ \|do| $R_{i,j} \assign \?from?[i]$ \\
5        \> \|return| $\?from?[i].\?value?$ \\
\|end|
\end{tabbing}
\end{minipage}
}
\caption{Construction~0}
\label{cons0}
\end{figure}

The timestamp function called for in lemma \ref{taglemma}
is built right into this construction.
Each operation starts by collecting value-timestamp pairs from all users.
In line 3 of either procedure, the operation picks a value and timestamp
for itself. It finishes after distributing this pair to all users.
It is not hard to see that the three conditions of lemma \ref{taglemma}
are satisfied for each complete proper history.
Integrity and Precedence are straightforward to check.
Uniqueness follows since timestamps of Write operations
of the single writer strictly increase (based on the observation that
each $R_{i,i}.\?tag?$ is nondecreasing). 

Restricting our unbounded timestamp multiwriter algorithm of \cite{vit86}, in 
the version of \cite{li89},
to the single writer case enabled us to tweak it to
have a new very useful property that is unique to the multireader case:
The greatest timestamp scanned by a reader is either the writer's
timestamp $\?from?[n].\?tag?$ or another reader's timestamp that is at most 1
larger. The tweaking consists in the fact that there is a definite
order in the scanning of the shared variables in line 2 of the Read
procedure: the writer's shared variable is scanned last 
(compare Section~\ref{sect.intuition}). 

\begin{lemma}\label{lem.C0}
The $\max$ selected in line 2 of the Read procedure satisfies
$\?from?[n].\?tag? \leq \?from?[\?max?].\?tag? \leq \?from?[n].\?tag? +1$.
\end{lemma}

\begin{proof}
At the time the writer's shared variable $R_{n,i}$ is scanned last
in line 2 of the $\?Read?(i)$ procedure, yielding $\?from?[n].\?tag?$,
the writer can have
started its next write, the ($\?from?[n].\?tag?+1$)th Write, but
no write after that---otherwise the $\?from?[n].\?tag?$
would already have been overwritten in $R_{n,i}$ by the
($\?from?[n].\?tag?+1$)th Write. Hence,
a timestamp scanned from another 
reader's shared variables $R_{j,i}$ ($j \neq n,i$)
can exceed the writer's timestamp by at most 1. 
\end{proof}

\section{Bounded}
\label{sect.bounded}

The only problem with Construction~0 is that the number of timestamps
is infinite. With a finite number of timestamps comes the necessity to
re-use timestamps and hence to distinguish old timestamps from new ones.
Our strategy will be as follows. We will stick very close
to construction-0 and only modify or expand certain lines
of code. The new bounded timestamps will consist
of two fields, like dominoes, the {\em tail} field and {\em head} field. 
\begin{definition}\label{def.tag}
A {\em bounded timestamp} is a pair $(t,h)$ where $t$ is the value of
the {\em tail} field and $h$ is the value of the {\em head} field.
Every field can contain a value $t$ with $0 \leq t \leq 4n+2$ or $t$ is the
distinguished initial, or bottom, value $\perp$. ($\perp$ is not a number
and is lower than any number.) 
We define the domination relation ``$<$'' on the bounded timestamps as follows:
$(t_1,h_1) < (t_0,h_0)$ if either $h_1 = t_0$ and $t_1 \neq h_0 \neq \perp$,
or $t_1,h_1 = \perp$ and $h_0 \neq \perp$.
\end{definition}

The matrix of shared variables stays the same, but now every
$R_{i,j}$ ($0 \leq i,j\leq n$) contains a $record$ 
consisting of a $value$ field, and a bounded timestamp consisting of
a $tail$ field and a $head$ field. The purpose is that process $j$
can see (read) whether process $i$ has executed a Read ($0 \leq i < n$)
or Write ($i=n$) with what timestamp. The readers use this information
to select an appropriate ``latest'' Write value to return in their Read. 
The shared variables $R_{i,n}$ for communication between 
readers $i$ ($0 \leq i \leq n-1$) and the writer $n$ contains
in addition a second such $record$. 
This second $record$ is used by the readers to
inform the writer of timestamps that are not obsolete,
in the sense that they
still exist in the local variables of the reader even though they
may have already disappeared from every shared variable in the system.
These timestamps may possibly be
written to a shared variable owned by the reader involved,
and hence cannot be recycled yet. 
The scan executed in line 1 of the Write
protocol gathers all timestamps written in both $record$s
of the $R_{i,n}$'s
($0 \leq i \leq n$). These constitute all
the timestamps that are not obsolete, and
hence the process can determine the $\leq 4n+2$ values occurring in
the fields (two per timestamp, one timestamp per $record$, for $2n+1$ $records$), 
and select a value that doesn't occur
(there are $4n+3$ values available exclusive of the bottom value $\perp$).
The initial state of the construction has all $record$s in the $R_{i,j}$'s
set to $(0, \perp , \perp)$. 

\begin{figure}
\fbox{%
\begin{minipage}[t]{3in}
\begin{tabbing}
\|type| \= \?shared? : \=  \|re| \= \kill
        \> \?shared? : \|record| \\
        \>             \>    \>\?value? : \abs\?type?  \\
        \>             \>    \>\?tail? \ : $\perp, 0 .. 4n+2$ \\
        \>             \>    \>\?head? \ : $\perp, 0 .. 4n+2$ \\
        \>             \> \|end|
\end{tabbing}
\end{minipage}
}

\medskip

\fbox{%
\begin{minipage}[t]{3in}
\begin{tabbing}
\|procedure| $\?Write?(n,v)$ \\
\|var| \= $j : 0 .. n$ \\
       \> $\?free? :$  $0 .. 4n+2$ \\
       \> \?temp? : \?shared? \\
       \> \?used? : $\|array|[0 .. 1 , 0 .. n]$ \|of| \?shared? \\
\|begin| \\
1        \> \|for| $j \assign 0 .. n-1$ \|do| $\?used?[0 .. 1, j]) \assign R_{j,n}$; $\?temp? \assign R_{n,n}$ \\
2        \> $\?free? \assign$ least positive integer not in \?tail? or \?head? fields of \?used? or \?temp? records  \\
3        \> $\?temp? \assign (v,\?used?[0,n].head,\?free?)$ \\
4        \> \|for| $j \assign 0 .. n$ \|do| $R_{n,j} \assign \?temp?$ \\
\|end|
\end{tabbing}
\end{minipage}
}

\medskip

\fbox{
\begin{minipage}[t]{3in}
\begin{tabbing}
\|procedure| $\?Read?(i)$ \\
\|var| \=    $j, \?max? $ \=   \= \\
       \> \?temp? : \?shared? \\
       \> $\?from? : \|array|[0 .. n]$ \|of| \?shared? \\
\|begin| \\
0.1        \> $\?from?[i] \assign R_{i,i}$; $\?temp? \assign R_{n,i}$ \\
0.2        \> $R_{i,n} \assign (\?from?[i], \?temp?)$ \\
1    	   \> \|for| $j = 0 .. n$ \|do| $\?from?[j] \assign R_{j,i}$ \\
2.1        \> \|if|  $\?from?[n] \neq \?temp?$ \|then|  \\
2.2       \> \>  $\?temp? \assign \?from?[n]$ ;
	   $R_{i,n}  \assign (\?from?[i] , \?temp?$) \\
2.3        \>   \> \|for| $j = 0 .. n$ \|do| $\?from?[j] \assign R_{j,i}$ \\
2.4       \>   \> \|if|  $\?from?[n] \neq \?temp?$ \|then|  \\
3.1         \>  \>  \> $ \; \; \; \?from?[i] \assign
(\?temp?.\?value?, \perp, \perp )$; \|goto| 4\\ 
2.5        \>   \|if| $\exists \?max?: 
              \?from?[\?max?].(\?tail?, \?head?)
           > \?from?[n]\cdot (\?tail?, \?head?)$ \|then|\\
3.2        \> \>  $\?from?[i]\assign \?from?[\?max?]$ \\ 
3.3       \> \|else| $\?from?[i]\assign \?from?[n]$ \\
4.1        \> \|for| $j \assign 0 .. n-1$ \|do| $R_{i,j} \assign \?from?[i]$;
         $R_{i,n} \assign (\?from?[i] \cdot \?temp?)$ \\
5        \> \|return| $\?from?[i].\?value?$ \\
\|end|
\end{tabbing}
\end{minipage}
}
\caption{Construction~1}
\label{cons1}
\end{figure}

The lines 
of Construction-1 are numbered
maintaining---or subnumbering---the corresponding
line numbers of Construction-0. The only real difference
in the Write protocols is line 2 (select new timestamp).
In the Read protocols the differences are
lines 2.x (determine latest---or appropriate---timestamp) and 
lines 3.x (assign selected timestamp to local variable
$\?from?[i]$),
and lines 0.x (start Read by reading writer's timestamp and writing
it back to writer). According to this scheme we obtain the 
out-of-order line-sequence
2.4, 3.1, 2.5, 3.2 because the instructions in lines 2 and 3
of Construction-0 are split and interleaved in
Construction-1.

\begin{lemma}\label{lem.free}
Line 2 of a Write always
selects a \?free? integer $h_1$ such that the timestamp
$(t_1,h_1)$ with new $\?head?$ $h_1$ and new $\?tail?$ $t_1$ ($\?head?$
of the timestamp of the directly preceding  Write), assigned in line 3
and written in line 4 of that Write, satisfies $(t_1,h_1) \not< (t_0,h_0)$
and $(t_1,h_1) \neq (t_0,h_0)$ for every timestamp $(t_0,h_0)$ either
scanned in line 1 of the Write, or presently occurring in a local variable
of a concurrent Read that will eventually be written to a 
shared variable in line 4 of that Read.
\end{lemma}

\begin{proof} 
{\em Intuition:}
In line 1 of a Write the shared variables with each reader,
and a redundant shared variable with itself,
are scanned in turn. In assigning a value
to $\?free?$ in line 2 the writer avoids all
values that occur in the $\?head?$ and $\?tail?$ 
fields of the shared variables. All local variables of a reader 
are re-assigned from shared variables in executing the Read procedure,
before they are written to shared variables. So the only
problem can be that $\?free?$ occurs in a local variable
of a concurrently active Read that has not yet written it
to a variable shared with the writer at the time that variable was scanned
by the Write.  If $\?free?$ exists
in a local variable $\?temp?$ this doesn't matter since the timestamp
concerned will not be used as a Read timestamp.
Hence, the only way in which
$\?free?$ can be assigned a value that concurrently exists in a local
variable of a Read which already has been used
or eventually can be used in selection line 2.5 of a Read,
and hence eventually can be part of an offending timestamp written
to a shared variable, is according to the following
scenario: A $\?Read?(i)$, say $R$, is
active at the time a Write, say $W_1$,
reads either one of their shared variables. This
$R$ will select or has already selected the offending 
timestamp $(t_0,h_0)$, originally written by a Write $W_0$ preceding $W_1$, 
but $R$ has not yet
written it 
to $R_{i,n+1}$ or $R_{i,n}$
by the times $W_1$ scans those variables.
Moreover, $W_1$ will in fact select a
timestamp $(t_1,h_1)$ with either $(t_1,h_1) < (t_0,h_0)$ 
or $(t_1,h_1) = (t_0,h_0)$. This can only happen if $\?free?$
in line 2 is $t_0$, or $h_0$ and the $t_0 = \?head?$ of the
timestamp of the Write immediately preceding $W_1$, respectively.  
Then, a later Read $R_0$ using in
its line 2.5 the $(t_1,h_1)$ timestamp written by $W_1$
and the $(t_0,h_0)$ timestamp written by $R$ concludes falsely
 that $W_1$ precedes $W_0$ or $W_1 = W_0$.
For this to happen, $R$
has to write $(t_0,h_0)$ in line 4, 
and has to assign it previously in one of lines
3.1, 3.2, or 3.3. In line 3.1 the timestamp assigned is 
the bottom timestamp $(\perp, \perp)$ which by definition
cannot dominate or equal a timestamp assigned by the writer. This
leaves assignement of the offending timestamp $(t_0,h_0)$
in line 3.2 or line 3.3. This also leads to a contradiction of which the
proof is deferred to a future paper using the I/O automata formalism.
\end{proof}

The crucial feature that makes the
bounded algorithm work is the equivalent of lemma~\ref{lem.C0}:

\begin{lemma}\label{lem.C1}
The timestamp selected in lines 2.x and
written in line 4 of a $\?Read?(i)$
is one of the following:

(i) The timestamp $(\perp, \perp)$ for a Write 
that is completely overlapped by the Read concerned; 

(ii) Another reader's timestamp scanned in line 1 (respectively, line 2.3) 
that is written 
by the next Write after the Write that wrote the
timestamp $\?from?[n].(\?tail?,\?head?)$ scanned in line 1.

(iii) Otherwise, the writer's
timestamp $\?from?[n].(\?tail?,\?head?)$ scanned in line 1.
\end{lemma}

\begin{proof} 
{\em Intuition:}
Only assignments in lines 3.1, 3.2, 3.3 can result in a write in line 4. 

(i) If line 3.1 is executed in a Read, then previously we scanned the writer's
timestamp in lines 0.1, 1, 2.3, and obtained three successive timestamps
without two successive ones being the same.
Hence the Write corresponding to the middle scan, of
line 1, is overlapped completely by the Read, and the Read can
be ordered directly after this Write, and this has no consequences
for the remaining ordering. This is reflected by using the timestamp
$(\perp, \perp)$ to be written by such a Read in line 4.

(ii) If line 3.2 is executed in a Read then the timestamp assigned is
a reader's timestamp scanned in line 1 or line 2.3.
The writer's timestamp used in the comparison line 2.5 is,
say, $(t,h)$.
According to the semantics of the 
``$<$'' sign in definition~\ref{def.tag},
only a reader's timestamp of the form $(h, \?free?)$ satisfies
the condition in line 2.5. The only timestamps in the system
are created by the writer.  
By Lemma~\ref{lem.free},
existence of
the $(h, \?free?)$ timestamp somewhere in the system at the time
of writing the current instance of timestamp $(t,h)$ would have prevented
the writer from writing $(t,h)$. Thus the writer must have
written the $(h, \?free?)$ timestamp after it wrote $(t,h)$.   
Using Lemma~\ref{lem.free} a second time,
the writer can write a timestamp with $h$ in the $\?tail?$ field
only at the very next Write after the Write that wrote a timestamp with
$h$ in the $\?head?$ field, since the $(t,h)$ timestamp is still
somewhere in a shared variable or to be written to a shared variable.

(iii) If line 3.3 is executed in a Read, then items (i) and (ii) were
not applicable and the timestamp assigned is the 
writer's timestamp scanned in line 1.    
\end{proof}

\begin{theorem}
Construction-1 is a wait-free implementation of an atomic 1-writer
$n$-reader register. It uses $(n+1)(n+2)-1$ atomic 1-writer 1-reader
$2 \log (4n+4)$ bits control shared variables. The Write  
scans  $2n+1$ of these variables and writes $n+1$ of them;
the Read scans $\leq 2n+3$ of these variables and writes $\leq n+3$
of them. 
\end{theorem}

\begin{proof}
{\bf Complexity:}
Since the program of Construction-1 contains no loops, it is
straightforward to verify that
every Read and Write executes the number of accesses to shared
variables, and the size of the shared variables,
 as in the statement of the theorem.

{\bf Wait-Freedom:} This follows directly from the upper bounds
on the complexity (scans and writes) of the shared variables in
the construction, and the fact that the program of Construction-1
is loop-free. 

{\bf Linearizability:}
Consider a complete proper history
(as defined before) $h$ of all Writes and only
those Read's that don't write the bottom timestamp $(\perp,\perp)$ in line 4.
Let $\ar_h$ be the partial order induced by the timing of the Reads and
Writes of $h$.
Lemma~\ref{lem.C1}, items (ii) and (iii), 
asserts that on this history $h$, Construction-0 and Construction-1 behave
identically in that the same Read reports the values of the same Write in
both constructions. 
Therefore, with respect to $h$, linearizability 
of Construction-1 follows from the linearizability of
Construction-0. 
Let $\ar_h^l$ be the 
linear order thus resulting from $\ar_h$.
The remaining Read's, 
those that write the bottom timestamp $(\perp, \perp)$ in line 4,
completely overlap a Write, lemma~\ref{lem.C1} item (i),
and report the value of that overlapped Write. Hence they can,
without violating linearizability, be inserted in the $\ar_h^l$-order
directly following the Write concerned.  
This shows that Construction-1 is linearizable. 
\end{proof}
 
{\bf Minimum Number of Global Control Bits:}
The same algorithm with only $O(n^2 )$ control bits overall
can be constructed as follows.
Each register owned by the writer contains $2n$ control bits,
and each register owned by a reader contains only 2 control bits.
The control bits are used to determine the domination relation
between readers and the writer. The Protocol stays the same,
only the decisions in the protocol are made according to
different format data. Since the decisions are isomorphic with
that of Protocol 1, the correctness of the new Protocol follows
by induction on the total atomic order of the operation executions
in each run by the correctness of Construction-1.

{\bf Minimum Number of Replicas of Stored Values:}
In the algorithm, each subregister ostensibly contains a
copy of the value to be written. This sums up to
$O(n^2 \log  V)$ bits, for the value ranging from 1 to $V$.
With the following scheme only the registers owned by the
writer contain the values. Each register owned by
the writer can contain two values. The two fields concerned
are used alternatingly. The writer starts its $t$th write with an extra write
(line 0)
to all registers it owns, writing the new value in field $t \pmod 2$.
In line 4 it writes to field $t \pmod 2$
(it marks this field as the last one written),
and finishes by setting
$t := (t+1) \pmod 2$.
The readers, on the other hand, now write
no values, only the timestamps. If the reader 
chooses the writer, it takes the value from the marked field;
if it chooses a reader, it takes the value from the unmarked field.
Since no observed reader can be more than one write
ahead of the actually observed
write, this is feasible while maintaining correctness.
This results in $O(n)$ replications of the value written resulting
in a total of $O(n \log V)$ value bits. This is clearly the 
optimal order since the writer needs to communicate the value to
everyone of the readers through a separate subregister. (Consider
a schedule where every reader but one has fallen asleep indefinitely.
Wait-freeness requires that the writer writes the value to a subregister
being read by the active reader.)

{\bf Acknowledgment:}
I thank Ming Li and Amos Israeli for initial interactions
about these ideas in the late eighties.
Other interests
prevented me from earlier publication. I also thank Grisha Chockler
for pointing out a problem with the previous formulation of 
Lemma~\ref{lem.free}, and meticulously reading
earlier drafts and providing useful comments.

\small{

}
\end{document}